\documentstyle[12pt]{article}

\def\void{}
\def\labelmark{}

\newenvironment{formula}[1]{\def\labelname{#1}
\ifx\void\labelname\def\junk{\begin{displaymath}}
\else\def\junk{\begin{equation}\label{\labelname}}\fi\junk}%
{\ifx\void\labelname\def\junk{\end{displaymath}}
\else\def\junk{\end{equation}}\fi\junk\labelmark\def\labelname{}}

{\ifx\void\labelname\def\junk{\end{array}\end{displaymath}}
\else\def\junk{\end{array}\right.\end{equation}}
\fi\junk\labelmark\def\labelname{}\def\junk{}
}

\newcommand{\beq}{\begin{formula}}
\newcommand{\eeq}{\end{formula}}
\newcommand{\beqv}{\begin{formula}{}}

\newcommand{\newsection}{
\setcounter{equation}{0}
\section}
\newcommand{\bea}{\begin{eqnarray}}
\newcommand{\eea}{\end{eqnarray}}
\newcommand{\rf}[1]{(\ref{#1})}
\newcommand{\g}{\gamma}
\renewcommand{\l}{\lambda}
\renewcommand{\b}{\beta}
\renewcommand{\a}{\alpha}

\newcommand{\k}{\kappa}
\newcommand{\m}{\mu}

\newcommand{\sg}{\sigma}

\newcommand{\oh}{\frac{1}{2}}
\newcommand{\dg}{\dagger}
\newcommand{\la}{\langle}
\newcommand{\ra}{\rangle}
\newcommand{\Tr}{{\rm tr}\;}

\newcommand{\cT}{{\cal T}}

\newcommand{\noi}{\noindent}

\begin{document}
\topmargin 0pt
\oddsidemargin 5mm
\headheight 0pt
\topskip 0mm

\addtolength{\baselineskip}{0.20\baselineskip}

\hfill NBI-HE-92-80

\hfill November 1992
\begin{center}

\vspace{36pt}
{\large \bf
The 3d Ising model represented as random surfaces}

\vspace{24pt}

{\sl J. Ambj\o rn }

\vspace{12pt}

The Niels Bohr Institute\\
Blegdamsvej 17, DK-2100 Copenhagen \O , Denmark\\

\vspace{18pt}

{\sl A. Sedrakyan}\\

\vspace{12pt}

Yerevan Phys. Institute \\
Alikhanian Brothers St. 2, 375 036 Yerevan \\
Armenia

\vspace{18pt}

{\sl G. Thorleifsson }

\vspace{12pt}

The Niels Bohr Institute\\
Blegdamsvej 17, DK-2100 Copenhagen \O , Denmark\\

\end{center}

\vfill

\begin{center}
{\bf Abstract}
\end{center}

\vspace{12pt}

\noi
We consider a random surface representation of the three-dimensional
Ising model.The model exhibit  scaling behaviour and a new critical index
$\k$ which relates $\g_{string}$ for the bosonic string to the exponent
$\a$ of the specific heat of the 3d Ising model is introduced. We try to
determine $\k$ by numerical simulations.

\vspace{24pt}

\vfill

\newpage

\newsection{Introduction}

The 2d Ising model formulated on the regular square lattice
has a well know random walk representation.
In fact we can write:
\beq{*1.1}
Z(\b) = \sum_{\sg} \exp (\b \sum_{\la nm \ra} \sg_n \sg_m) \sim
\sum_{P} \Phi (P)\; \frac{1}{L(P)} \exp (-\m(\b) L (P)),
\eeq
\beq{*1.2}
\m(\b) = -\ln \tanh \b, ~~~~~~\Phi(P) = (-1)^{n(P)} .
\eeq
In these formulas $P$ is a closed (not necessarily connected) path without
back-tracking\footnote{ Back-tracking means in this context that two
successive steps in the lattice path shares the same link. For surfaces
the generalization is that two neighbouring plaquettes do no lie on
top of each other.}
on the lattice, $L(P)$ the length of the path (i.e. the number of links
constituting  the path)
and $n(P)$  the number of times the path $P$ self-intersects.
The sum of paths exponentiate as a  sum over connected closed paths and
the free energy of the 2d Ising model:
\beq{*1.3}
F(\b) \sim \ln Z(\b) \sim
\sum_{{\rm connected}~ P} \Phi(P) \frac{1}{L(P)}
\exp \left(-\m(\b) L (P)\right).
\eeq
It is known that the 2d Ising model has a fermionic representation,
a fact closely related to the sign factor $\Phi (P)$, and that the
continuum field theory associated with the second order transition
of the 2d Ising model is that of a free Majorana spinor.

If we consider the 3d Ising model defined on a regular cubic lattice
it has a random surface representation\cite{fss} which can been viewed as the
analogue of the random walk representation \rf{*1.1}-\rf{*1.2}.
It can be written as
\beq{*1.4}
Z(\b) = \sum_{\sg} \exp (\b \sum_{\la nm \ra} \sg_n \sg_m )
\sim
\sum_{S} \Phi (S)\; \frac{1}{C(S)} \exp \left(- \m(\b) A (S) \right)  .
\eeq
\beq{*1.5}
\m(\b) = - \ln \tanh \b,~~~~~\Phi(S) = (-1)^{l(S)},
\eeq
where $S$ denotes a closed lattice surface (without back-tracking)
built from plaquettes, $C(S)$ a symmetry factor for the surface
\cite{itzykson}, $A(S)$  the lattice area  of $S$
(i.e. the number of plaquettes constituting the surface $S$)
while $l(S)$ is the number of links where the surface intersects
itself. $(-1)^{l(S)}$ is obviously a surface analogy  of the Kac-Ward
factor $(-1)^{n(P)}$ for the random walk. Again the sum over
surfaces exponentiates into a sum over connected closed
surfaces\cite{itzykson}:
\beq{*1.6}
F(\b) \sim \ln Z(\b) \sim
\sum_{{\rm connected}~ S} \Phi (S) \;
\frac{1}{C(S)} \exp \left(- \m(\b) A (S) \right) .
\eeq

Although it is by no means clear how to take the continuum limit of this
lattice surface theory, due to the oscillating sign factor $\Phi (S)$,
one would nevertheless think  that the second order transition of the
3d Ising model should result in continuum field theory, and that this
continuum theory should allow a representation like \rf{*1.6}, only
with continuum surfaces instead of lattice surfaces.
For a closed path $P$ in two dimensions it is possible to obtain a simple
geometrical expression of the sign factor $\Phi (P)$:
\beq{*1.7}
\Phi (P) = \exp \left( \oh i \oint   e_i de_i \right),
\eeq
where $e_i(\xi)$ is the tangent vector on the path $P$,
but an analogous expression is not known for $\Phi(S)$ for
a surface $S$ immersed in $R^3$. One can  nevertheless
start on the lattice, introduce a fermionic representation
of $\Phi (S)$ and take the formal continuum limit of the lattice
fermionic action \cite{ks,s} (see also \cite{orland} for related
considerations). In this way the continuum limit
becomes a string theory where the action is sum of an area term
and the induced Dirac action plus an anomaly term
on the world surface immersed in $R^3$ and
we can formally view $\Phi (S)$ as a result of first integrating
over the fermions. If we accept this point of view, it makes
sense to talk about a topological expansion of the surface theory,
which at first glance might appear somewhat surprising since the
original  high temperature expansion over lattice surfaces
involves crucial cancellations between surfaces of different
topologies and cancellations between orientable and non-orientable
surfaces. If we {\it assume} that such a topological expansion is possible
close to the critical point of the Ising model\footnote{It should be stressed
that it seems difficult not to have a topological
expansion, possibly involving both orientable and non-orientable surfaces,
if there really exists a continuum fermionic surface representation at
the critical point. Viewed in this way the assumption of a topological
expansion is more or less equivalent with the assumption of a continuum
surface representation. We feel there are good reasons to expect the
existence of such a continuum surface representation \cite{ks,s}.}
we can discuss  scaling properties of $\Phi (S)$ which are consistent
with this assumption and the known critical behaviour of the Ising model.

Since the 3d Ising model has a second order phase transition at a certain
value $\b_c$ we have
\beq{*1.8}
F(\b) = {\rm r.t}(\b) + c (\b_c-\b)^{2-\a} + {\rm l.s} (\b)
\eeq
where r.t$(\b)$ stands for regular terms, l.s$(\b)$ stands for less singular
terms, while the critical exponent $\a$ is known from high temperature
expansions and renormalization group study to be $\a \approx 0.11$
or $0.125$ \cite{gz,h}.
How can such an expression match the left hand side of \rf{*1.6}?
If we ignored sign factor $\Phi (S)$ we would have an ordinary
bosonic string theory\footnote{In the following
we assume the world sheet variables
are somehow regularized, either by the original cubic lattice or by
dynamical triangulations.}. The entropy factor of bosonic surfaces
of a fixed topology has (at least for $c <1$) the form
\beq{*1.9}
Z(A) \sim A^{\g-3} e^{\m_0 A}.
\eeq
This leads to the well known singularity structure of the bosonic string:
\beq{*1.10}
Z(\m) = \sum_{A} Z(A) e^{-\m A} \sim
{\rm r.t}(\m) + c (\m-\m_0)^{2-\g} + {\rm l.s} (\m)
\eeq
which seems unrelated to \rf{*1.8}, since $\m_0 > \m (\b_c)$.
It is however intuitively  clear that  $\la \Phi(S) \ra$ is small and
it is reasonable to assume it has the following asymptotic
behaviour as a function of the area $\bar{A}$:
\beq{*1.11}
\Phi_{\bar{A}} \equiv
\frac{\sum_{\{S\; |\; A(S)=\bar{A}\}}
\Phi (S)e^{-\m_0 A(S)}}{\sum_{\{S\; |\; A(S)=\bar{A}\}}
e^{-\m_0 A(S)}} \sim {\bar{A}}^\k\; e^{-\m_1 \bar{A}}.
\eeq
Under this assumption the rhs of \rf{*1.6} can be written   as
\beq{*1.12}
\sum_{A} Z(A)\;\Phi_A \; e^{-\m A} \sim
{\rm r.t}(\m) + c (\m-(\m_0-\m_1))^{2-\g-\k} + {\rm l.s} (\m)
\eeq
and comparing with \rf{*1.8} we reach the conclusion that
\beq{*1.13}
\a = \g + \k,~~~~~~\m_0-\m_1 = \m(\b_c).
\eeq
\vspace{12pt}

At this point it should be made clear that the same arguments can be used
in the case of  the ordinary bosonic random walk,
$\Phi(P)$ and the 2d Ising model. It
can be shown that we indeed have a relation similar to \rf{*1.11}:
\beq{*1.11a}
\Phi_{\bar{L}}\equiv
\frac{\sum_{\{P | L(P) = \bar{L}\}} \Phi(P) e^{-\hat{\m}_0 L(P)}}{
\sum_{\{P | L(P) = \bar{L}\}}  e^{-\hat{\m}_0 L(P)}}
\sim {\bar{L}}^{\hat{\k}} \;e^{-\hat{\m}_1 \bar{L}}.
\eeq
The exact values of $\hat{\m}_0$ and $\hat{\m}_1$
depends on the discretized models
used for the bosonic random walk. The result \rf{*1.11a} is implicitly in the
formalism developed for discretized fermionic walks in \cite{adj}, where
the phase factor \rf{*1.7} is represented as a product of Hilbert-Schmidt
operators on $S^1$. The largest eigenvalue of that operator
dominates in the scaling limit
and one can prove  that $\hat{\k}=-1$  in agreement
with \rf{*1.13} since $\g=1$ for a closed random walk in two dimensions
 and $\a=0$
(logarithmic divergence) for  the 2d Ising model.

\vspace{12pt}

In the case of random surfaces a formula like \rf{*1.13}
immediately leads to the interesting question of
topology dependence: $\m_0$ is independent of topology and
it is reasonable to assume that the same is true for $\m_1$.
This assumption is needed for
$\m_0 -\m_1 = \m(\b_c)$ to make sense, since
$\b_c$ has no a priori relation with the topology.
We know on the other hand that $\g$ depends on topology in a linear
way. For central charge less than or equal to one we have as a
function of genus $g$ of the surface:
\beq{*1.14}
\g_g = \g_0 + K(c)g, ~~~~K(c) >0.
\eeq
$K(c)$ is not known for $c > 1$. In case the
surfaces for $c >1$ are dominated by so-called branched polymers we
know that $K(c) < 0$ for sufficiently large $c$ \cite{adj1,paolo}.
If $K(c) < 0$ it is plausible that
surfaces of spherical topology will dominate the critical behaviour,
but  this dominance  could in principle be eliminated by a
$g$-dependence of $\k$. If that is the case the simplest scenario would
be one where all genus contributions gave rise to the same $\a$-dependence.
In both case we should be able to test $\a = \g+ \k$ by considering
spherical topology.

In sect. 2 we define the discretized model we are going to use in our
attempt to verify $\a = \g +\k$. The results of numerical simulations
are discussed in sect. 3.

\newsection{The model}

Let us recall the  construction of the sign factor $\Phi(S)$ for the regular
lattice surface \cite{ks,russians}.
We shall cover the lattice surface by a system of closed non-self-intersecting
curves, which cross each link ones. This type of covering by curves can be
obtained by drawing two parallel lines connecting the midpoints of neighbouring
links on each plaquette of the surface. There are two ways of drawing these
lines on each plaquette (see fig.1 in ref. \cite{ks}) and we have a class of
$2^{M}$ coverings (M is the total number of plaquettes of the surface). Each
covering consists of a set of closed non-self-intersecting curves
$C_1,...,C_m$.
Let $n (C)$ be the number of times the curve $C$
crosses the lines of self-intersection for the surface $S$ immersed in the
cubic lattice.
Assume we can find a function $\Phi (C)$ such that
\beq{*2.1}
\Phi (C) = (-1)^{n(C)}.
\eeq
It follows that
\beq{*2.2}
\Phi(S) = (-1)^{l(S)} = \prod_k (-1)^{n(C_k)} = \prod_{k} \Phi (C_k)
\eeq
where the product is over all curves of the chosen curve system.
We can define a function $\Phi (C)$ with the above mentioned properties by
 \cite{ks} (this definition makes sense even for a continuum surface):
\beq{*2.3}
\Phi ( C) = \oh \Tr {\rm P} \exp \left( \oint \; R^{-1} dR \right).
\eeq
The  matrix $R$ is defined as follows: Consider three
orthonormal vectors $e_i^\a$, $\a=1,2,3$, attached to each point of the curve
$C$ : $e^1_i$ is the tangent vector, $e^2_i$ is normal to $e^1_i$ in the
tangent plan of the surface and $e^3_i$ is normal to the surface.
Then  $R$ rotates this orthonormal
system into a given fixed orthonormal system. If the rotation matrix
$R$ is in the fundamental representation and $\sg_i$ denote the Pauli matrices
this can be written as follows:
\beq{*2.4}
R  \tau_\a R^{-1} = \sg_\a,~~~~~\tau_\a \equiv e^\a_i \sg_i
\eeq
It can be shown that $\Phi(C)= 1$ if $C$ does not cross a line of
self-intersection
 and --1 if it crosses it once. In addition $\Phi(C)$ is invariant under smooth
 deformations of $C$ since it takes discrete values. Only when the deformation
 of $C$ reaches singular points  of the immersed surface, where $R$ is not
 defined, will it be able to change its value. The singular points
 are the possible endpoints
 of lines of self-intersection of the immersed surface and these points
 are so-called Whitney singularities, which are related to
 $\pi_1 (SO(3))$, i.e. elements like \rf{*2.3}. In the case of piecewise
 linear surfaces (either the cubic lattice considered above or the triangulated
 random surfaces immersed in $R^3$, to be considered below) we will assume
 that the curve $C$ is a straight line on the flat part of the surfaces.
Then the integral \rf{*2.3} will only get contributions when the curve
 crosses from one flat piece to a neighbouring one. Let us denote the
 successive flat pieces of surface encountered by the curve by $S_n$,
 the corresponding orthonormal frames  by $e^\a(n)$ and the the rotation
 matrices related to $e^\a(n)$ by $R(n)$.
 With this notation \rf{*2.3} can then be written as
 \beq{*2.5}
 \Phi(C) = \oh \Tr \prod_n
\frac{1+ \tau_\a (n) \tau_\a (n+1)}{\sqrt{1+e_i^\a (n)e_i^\a(n+1)}}
\eeq
 As was shown in ref.\cite{ks} the sign factor $\Phi(S)$ does not depend on
the specific choice of one of $2^{M}$ coverings.
One can choose the covering such that we have a maximum
number of closed non-connected curves.
These curves will then surround half of
the lattice vertices separating them into two classes, say ``+''
and ``--''
vertices (see \cite{russians}). Then the sign factor $\Phi(S)$ can be said
to be located at one type of the vertices, say the ``+'' vertices, and by
construction we have
 \beq{*2.5a}
 \Phi(S) = (-1)^{n_{+}}
 \eeq
 where $n_{+}$ is the number of Whitney singularities, located at the ``+''
vertices \footnote{Note that on piecewise linear surfaces the Whitney
singularities have to be located on the vertices (the 0-simplexes) of the
surface.}

Rather than applying the formula \rf{*2.5} or \rf{*2.5a}
 to the cubic lattice surfaces we
 will assume that there is indeed a continuum surface theory of the
 3d Ising model at the critical point, and that
 we are free to use any sensible approximation to the
 continuum surface theory, as long as it admits us to take the correct scaling
 limit. From this point of view it seems reasonable to use  triangulated
 piecewise linear surfaces immersed in three dimensions as our statistical
 ensemble of surfaces. Their critical properties coincide with those of the
 bosonic string in target space dimensions $d < 1$, where the theory can
 also be solved analytically.  We have however to choose an appropriate
 curve system on the piecewise linear surface and for that purpose
 the generic triangulation of  a surface is not useful. Rather we consider
 the class of triangulations obtained by gluing squares together as
 indicated in fig. 1. Each square consists of two triangles, one of the
 diagonals of the square being the common link of the two triangles.
 The four vertices are divided into two groups: ``--''vertices connected
 by the diagonal and ``+''vertices separated by the diagonal. The rule
 for gluing together links of different
 squares to form a (triangulated) surface is that  ``+''vertices should
 be glued to ``+''vertices and ``--''vertices to ``--''vertices
 (see fig. 1). For this
 class of triangulated surfaces one can now choose a convenient curve
 system which separates the ``+'' and the ``--'' vertices. For each square
 two parts of different curves run parallel to the diagonal connecting
 the ``--''vertices. Stated differently we can say that the closed
 curves on the surface surround all ``+''vertices (see fig. 1).
 From this point of view the curve system is identical to one described
above on the cubic lattice,
 but of course the connectivity of the
 triangulated surfaces are such that they cannot in general be
 mapped into a cubic lattice.
 The class of triangulations generated this way can be
viewed in another way: The triangles surrounding ``+''vertices
form polygons, where the vertices on the boundaries are ``--''vertices,
and precisely one curve from the curve system runs around the ``+''vertex
inside this polygon. The surface is now constructed by gluing together
these polygons to form a connected closed surface.

As this class of triangulations is different from the one usually
used in discretized 2d quantum gravity one could worry if they
belong to the same universality class, which is clearly what we
want. One can easily prove this. It is possible to write down
a matrix model which precisely generates the triangulations we
are considering:
\beq{*2.10}
Z(\l) = \int d\phi  d\phi^*  dA \;
e^{- \Tr \phi^\dg \phi - \oh \Tr A^2 + \l \Tr \phi^\dg \phi A} .
\eeq
In \rf{*2.10} $\phi$ denotes a complex $N\times N$ matrix, $A$
an Hermitian $N\times N$ matrix and to $\Tr \phi^\dg \phi A$ we can
associate an oriented  triangle with one ``+''vertex and two ``--''vertices.
The link connecting the ``--''vertices corresponds to the $A$ matrix.
If we first integrate over the
Hermitian matrices we reproduce precisely the complex matrix
model with quartic interactions which is known to have the same
$\g_{string}$ as the corresponding Hermitian  matrix model
\cite{am,morris,ajm}.
The integration over the Hermitian matrices corresponds  creating the
squares of fig. 1 by identifying the ``--'' vertices of two  triangle.
Alternatively the integration over the complex  matrices will leave
us with an Hermitian matrix model with potential a $\log(1-A)$. This
corresponds to  gluing together  polygons as mentioned above\footnote{Note that
we have by these arguments shown that to any order in $1/N$
the $m=2$ Hermitian matrix
model has the same critical behaviour as the $m=2$ complex matrix
model, a well know result which was however
not entirely trivial to prove.}.

An immersion in $R^3$ of a given (abstract) triangulation $T$  of
a surface is a map from the vertices $i \in T$ into $x_i \in R^3$, such that
 links $\la ij \ra$ are mapped into straight lines connecting
 $x_i$ and $x_j$ and
 triangles $\la ijk \ra$ are identified with the triangles spanned by
 $x_i,x_j,x_k$. This results in a piecewise flat surface $S(\{x_i,T\})$.
 The curve system defined above is by the same procedure mapped into
 piecewise linear curves on $S$ and
we can use \rf{*2.2} and \rf{*2.5} to construct the sign factor
$\Phi(S(\{x_i,T\})$.
The definition of the model of random surfaces is as follows:
\beq{*2.11}
F(\m) = \sum_{T \in \cT} \frac{e^{-\m N_T}}{C(T)}\; \;
\int \prod_{i \in T} dx_i\; \delta^3( \sum_i x_i)\;\;
e^{ -\sum_{\la ij \ra} (x_i -x_j)^2} \; \Phi(S\{x_i,T\} )
\eeq
where $N_T$ denotes the number of triangles,
$C(T)$ is a symmetry factor for the
triangulation $T$ and $\cT$ denotes the class of triangulations defined above.
In principle we can include a summation over different topologies
in the sum over
triangulations, i.e. in the class $\cT$, but as discussed above we will
here try to test the conjecture that surfaces of spherical topology will
be important in the scaling limit. This we do  by comparing the
critical exponent $\a$ of the 3d Ising model with $\g$ and $\k$
extracted from $F(\m)$ defined \rf{*2.11}, but
with $\cT$ restricted to spherical topology.

The free bosonic surface theory corresponding to \rf{*2.11} is given
by\cite{adf,david,kkm,adfo}
\beq{*3.1}
F_0(\m) = \sum_{T \in \cT} \frac{e^{-\m N_T}}{C(T)}\; \;
\int \prod_{i \in T} dx_i\; \delta^3( \sum_i x_i)\;\;
e^{ -\sum_{\la ij \ra} (x_i -x_j)^2}.
\eeq
It is in  this ensemble  we will try to calculate the expectation value
$\la \Phi (S) \ra_0$. We expect an exponential fall off like
in \rf{*1.11}, only with $N_T$ instead of the area $A$.

\newsection{Numerical method and results}

Unfortunately
we have  at this stage to rely on numerical methods. One such
method is Monte Carlo simulations. It has been applied extensively
to simulations of the bosonic string defined by \rf{*3.1}. Two types
of simulations have been used, a ``canonical'' one where the number
of triangles $N_T$ are kept fixed, and a ``grand canonical'' one where
$N_T$ is allowed to vary\cite{adfo,jkp,adf1}.
The standard canonical move, the so-called
``flip'' of a link, is not available for us since we consider a restricted
class of triangulations $\cT$. In fact, since we are also interested
in a determination of $\g_{string}$, we have chosen the grand canonical
updating scheme, which applies to our class of triangulations
with a small modification of the standard moves. Each ``+''vertex
is surrounded by ``--''vertices and each ``--''vertex is surrounded
by an alternating sequence of ``+'' and ``--''vertices. We have shown
the corresponding moves for inserting and deleting ``+'' and ``-'' vertices,
respectively, in fig. 2. We have checked numerically that
this algorithm indeed leads to the correct value
$\g_{string}=-1/2$ in the case
where the dimension of spacetime is $d=0$, but it should be mentioned
that the extraction of the correct value was more difficult than if
we used the standard ensemble of triangulations, i.e. all triangulations
with spherical topology and  length of loops of links larger than or equal
three.

Let us now turn to the 3d case. There has been a number of
attempts to determine $\g_{string}$ numerically \cite{adf1,jkp,djkp,af,abk}.
The conclusion from these simulations is that $-0.2 < \g_{string} < 0.2$.
for $d=3$.
We will not improve
these simulations, since we will be working with quite small
surfaces and the only difference compared to these earlier
investigations will be that our class of triangulations is different,
but (at least in zero dimensions) belongs to the same universality
class. Our results for $\g_{string}$ are compatible with the earlier
measurements, indicating that also in three dimensions does our class
of triangulations give the same results as the unrestricted class of
triangulations.

The quantity which has our main interest is
$\la \Phi_{N_T} (S) \ra_0$, where $N_T$ is the number of triangles which
constitutes the surface. $ \Phi_{N_T} (S)$ is always $\pm 1$
but will fluctuate
wildly, both when we change triangulations and when we update
the coordinates $x_i$, and in accordance which our expectations
$\la \Phi_N (S) \ra_0$ falls off exponentially with the number of
triangles. In practise this means that we  have not been able to use surfaces
with $N > 34$. These are quite small surfaces, but hopefully large
enough\footnote{ At this point it is worth to recall the history of the
determination of $\g_{string}$ by numerical methods. Although the first
simulations used very small surfaces the qualitative features extracted
from these simulations are essentially unchanged today, years and
thousands of CPU hours later.}
to reveal the qualitative behaviour of the critical
exponent $\k$ in $\la \Phi_N (S) \ra_0 \sim N^\k \exp(-\m_1 N)$.

In table 1 and fig. 3 we have shown the result of approximately $10^9$ sweeps
over surfaces with $N$ between 6 and 40.

\begin{table}[t]
\begin{center}
\begin{tabular}{|l|l|}  \hline
  $N_T $  & \multicolumn{1}{c|}{ $\la \Phi_{N_T}(S) \ra_0$}  \\ \hline
  6       & $0.38570 \pm 0.00018$   \\
  8       & $0.29515 \pm 0.00013$   \\
  10      & $0.17897 \pm 0.00012$   \\
  12      & $0.09469 \pm 0.00012$   \\
  14      & $0.05903 \pm 0.00013$   \\
  16      & $0.03221 \pm 0.00014$   \\
  18      & $0.01882 \pm 0.00017$   \\
  20      & $0.01127 \pm 0.00011$   \\
  22      & $0.00606 \pm 0.00017$   \\
  24      & $0.00309 \pm 0.00018$   \\
  26      & $0.00190 \pm 0.00014$   \\
  28      & $0.00113 \pm 0.00017$   \\
  30      & $0.00062 \pm 0.00033$   \\
  32      & $0.00024 \pm 0.00019$   \\
  34      & $0.00040 \pm 0.00019$   \\
  36      & $0.00012 \pm 0.00020$   \\ \hline
\end{tabular}
\end{center}
\caption{measurement of $\la \Phi_{N_T}(S) \ra_0$}
\end{table}

The best value one can extract from the data seems to be $\k=0  \pm 1$.
The determination of $\k$ is rather  poor and can only be improved by going
to larger surfaces, which is impossible because of the exponential fall off
of the sign factor with the size of the surface. We nevertheless find
the results encouraging in the sense that they are compatible which
the relation $\g_{string}+ \k = \a$. It would obviously be very desirable
to be able to find observables which have  a better behaviour
than $\la \Phi_N \ra_0$ for large $N$ and which still relate critical
exponents of the Ising model to those of random surfaces.
The situation is the same as for the measurements
of Wilson loops in lattice gauge theories. The expectation value
of the Wilson loops will always be exponentially suppressed because
we make the importance sampling in the pure gauge theory which knows nothing
about the static quarks.
In the same way we here make the importance
sampling  of pure bosonic string configurations
without knowledge about the fermionic string which
describes the Ising model.
If better  observables
could be found it would be possible to go to larger surfaces and thereby
get more reliable critical exponents and in addition check for the
contributions of different topologies and orientable versus non-orientable
surfaces.

\vspace{12pt}

Our results indicate that surfaces of spherical topology satisfy the
scaling relation $\g_{string}+\k=\a$
near the critical point in a random surface representation of the 3d
Ising model. We are fully aware that our numerical
results are such that ``indicate'' is the appropriate  word to use.
As discussed above even an improved numerical verification of this
relation would not {\it prove} that only surfaces of spherical topology
are important in the scaling limit, but it puts emphasis on such a
possibility. In principle one could dream of settling the problem
of dominance of spherical topology by numerical simulations since it is
possible to perform simulations with surfaces of topology different from
that of the sphere and also to include non-orientable surfaces. However,
we need better observables, which do not fall off so fast with the size of
the surfaces, if we want to be able to make quantitative statements.

But apart from attempts to improve the numerical simulations
we hope that our point of view might simulate theoretical attempts to
understand better the random surface representations of the 3d Ising model.

\vspace{24pt}

\noindent
{\bf Acknowledgement} We thank Peter Orland for reading the manuscript
and for valuable comments.

\vspace{24pt}

\addtolength{\baselineskip}{-0.20\baselineskip}

%\newpage

%\vspace{36pt}
\newpage

\noindent {\large \bf Figure Captions}

\vspace{12pt}

\begin{itemize}

\item[Fig.1] The matrix model representation of our restricted
class of triangulations. Vertices with open circles represent the
``--'' vertices defined in the text, while vertices with black
circles represent the ``+'' vertices. The dotted diagonals symbolize
the integration over the Hermitean matrix $A_{\g\b}$, while the dashed
line shows the curve $C_v$ surrounding the ``+'' vertex $v$.

\item[Fig.2] The two classes of moves for inserting and deleting
``-'' vertices (open circles) and ``+'' vertices (black circles).

\item[Fig.3] A graphical representation of the data of table 1.
We expect for large $N$: $$\log [\la \Phi_N \ra/\la \Phi_{N-1} \ra] \sim
-\mu_1
+ \k \log[N/(N-1)].$$

\end{itemize}

\end{document}